\documentclass[12pt]{article}

\begin{document}

\begin{titlepage}

\title{Hermitian vs PT-Symmetric  Scalar Yukawa Model}

\author{  {\bf V. E. Rochev}\footnote{E-mail address: rochev@ihep.ru}
\\  State Research Center of the Russian Federation\\ 
 ``Institute for High Energy Physics''\\
of National Research Centre  ``Kurchatov Institute``,\\ Protvino, Russia}
\date{}

\end{titlepage}
\maketitle
\begin{abstract} 

A comparative analysis of
a model of complex scalar field $\phi$
and real scalar field $\chi$ with  interaction
$g\phi^*\phi\chi$
for the real and purely imaginary
 values of  coupling $g$  in perturbative and non-perturbative
regions. In contrast to the usual Hermitian  version (real $g$), which is
asymptotically free and energetically unstable, the non-Hermitian $PT$--symmetric
 theory (imaginary $g$) is energetically stable and not asymptotically free.
The non-perturbative approach based on  Schwinger--Dyson equations
reveals new
 interesting feature of the non-Hermitian model. While in the Hermitian version of theory the 
phion propagator 
 has the non-physical non-isolated singularity in the Euclidean 
 region of momenta, the non-Hermitian theory  
substantially free of this drawback, as the singulatity  moves in the pseudo-Euclidean
region.

\end{abstract}


\section{Introduction}
Non-Hermitian $PT$--symmetric quantum models, open at the end of the last century, \cite{Bender98},
currently have a wide use in various fields of physics (see. review of \cite {Bender15}
and references therein). For a quantum field theory introduction into circulation the non-Hermitian
$PT$--symmetric models 
interesting enough as an extension of a narrow class of Hermitian models with acceptable
physical and mathematical properties (such as stability, unitary, renormalisability, etc.)
and
opens new possibilities for describing the properties of high-energy particles.

In  works of Bender et al. \cite{Bender12}, \cite{Bender13} the $PT$--symmetric model
of a scalar field 
with the interaction $\phi^3$ has been investigated.
As  has long been known (see. \cite{woo}), the Hermitian version of
this model is asymptotically free, but the unstability of the cubic interaction
 leads to the fact that models of this type previously considered exclusively
as a methodical examples  (see, for example,
 \cite{Collins}). For the non-Hermitian
version of this model with imaginary coupling, however, the main argument of
the unstability
 -- the cubic potential is
 unbounded below  --   becomes invalid since 
complex numbers are not an ordered set.
 Moreover, in work \cite{Bender12}, the arguments are given in favor of energy
stability of the non-Hermitian model with cubic interaction.   
The  analysis of Bender et al (see \cite{Bender12},  \cite{Bender13}) 
indicates that the $ig\phi^3_6$ theory is like a $g\phi^4_4$ theory:
it is energetically stable, renormalized and has the  trivial-type ultraviolet behavior, i.e.,
compared with the conventional $\phi^3_6$ model, the $PT$--symmetric $ig\phi^3_6$ theory exhibit new 
 interesting properties.

In this paper  we study the scalar Yukawa model, i.e,  a model of a complex scalar
field $\phi$ (phion) and a real field $\chi$ (chion) with the 
interaction $g\phi^*\phi\chi$.
This model is used in nuclear
physics as a simplified version of the Yukawa model without spin
degrees of freedom, as well as an effective model of
the interaction of scalar quarks \cite{Guasch}, \cite{abreu}.
If the coupling constant $g$ takes purely imaginary values  and the field $\chi$
is a pseudoscalar, such a model is $PT$ -- symmetric.
As expected, this
model is a very similar to the $\phi^3$ theory.
All arguments of Bender et al  (see \cite{Bender12}) concerning the  unstability of the Hermitian  theory and
stability of non-Hermitian $PT$ -- symmetric theory fully extended to the scalar Yukawa model.
An additional argument is the consideration in the spirit of \cite{Cornwall} a zero-dimensional
version of the theory. The partition function
$$
G(g)=\int
D(\phi,\phi^*,\chi)\exp\bigg\{\int dx\, {\cal L}(x)
\bigg\}
$$
in a zero-dimensional space becomes  the usual improper integral
$$
G(g)=\int_{-\infty}^{\infty}
d\phi d\phi^* d\chi\exp\bigg\{-\phi^*\phi
-\frac{1}{2}\chi^2+g\phi^*\phi\chi\bigg\}=
\sqrt{2\pi}\int_{-\infty}^{\infty}
d\phi d\phi^* \exp\bigg\{-\phi^*\phi
+\frac{g^2}{2}(\phi^*\phi)^2
\bigg\},
$$
which converges for   $g^2<0$ (non-Hermitian  theory) and diverges for $g^2> 0$ (Hermitian  theory).

In the coupling-constant perturbation theory,
 this model also has a very similar to the  $\phi^3$ theory.
Section 2 briefly presents the results of the coupling-constant
 perturbation theory and based on the perturbation theory
renormalization-group analysis for this model. As well as the $\phi^3_6 $  theory
the Hermitian scalar
Yukawa model in a six-dimensional space is asymptotically free. The Non-Hermitian scalar
Yukawa model in the $d = 6- \epsilon$ has, besides the Gaussian fixed point, also the non-Gaussian
fixed point  of Wilson - Fisher type. At $d = 6$  the non-Hermitian scalar Yukawa model, as well as
$\phi^ 3_6$  theory is ultraviolet unstable, and to describe the ultraviolet region
we need to go beyond the perturbation theory.

Section 3 presents  an attempt to go beyond the coupling-constant perturbation theory. The formalism of
bifocal source is used to built a non-perturbative expansion of the system  of the Schwinger-Dyson equations, 
 and 
 equation for the phion propagator in the leading approximation of this expansion is investigated.
A  remarkable property is established: for the Hermitian theory
 the  phion propagator  has a non-isolated singularity
 in the Euclidean region of momenta while for the Hermitian theory this singularity  (an origin of
a cut) moves in a pseudo-Euclidean region,
 i.e., from the point of view of the analytic properties of the propagator the
non-Hermitian theory is preferable.

\section{Perturbation Theory and Renormalization Group}

\subsection{Perturbation Theory}

We consider the model of interaction of a
complex scalar field $\phi$ (phion) and a real scalar field $\chi$
(chion) with the Lagrangian
\begin{equation}
  {\cal L}= -\partial_\mu\phi^*\partial_\mu\phi-m^2_0\phi^*\phi-
\frac{1}{2}(\partial_\mu\chi)^2-\frac{\mu^2_0}{2}\chi^2+g\phi^*\phi\chi
\label{Lagr}
\end{equation}
 in a $d$--dimensional Euclidean space 
$(x\in E_d)$ near $d=6$. At $d=6$ the coupling  $g$ is  dimensionless, 
and the theory   contains  ultraviolet divergences which can be
eliminated with a standard recipe by the renormalization of fields and vacuum
expectations (Green's functions).

The perturbation theory on the renormalized coupling constant $g$
 gives us the following
expressions for the renormalized 1PI functions:\\ 
Propagators of the phion
\begin{equation}
 \Delta^{-1}(p)=m^2+p^2-g^2L(p^2, \mu^2, m^2)+\delta m^2+p^2\delta z_1+O(g^4)
\end{equation}
and of the chion
\begin{equation}
 D^{-1}(k)=\mu^2+k^2-g^2L(k^2, m^2, m^2)+\delta \mu^2+k^2\delta z_2+O(g^4).
\end{equation}
A vertex:
\begin{equation}
 \Gamma(p_x, p_y)=g+g^3\Lambda(p_x, p_y)+\delta g+O(g^5)
\end{equation}
Here $m, \mu $ are the renormalized masses of the phion and the chion.
$\delta m^2,  \delta z_1, \delta \mu^2, \delta z_2$  are  counter-terms of the renormalization of
the masses and fields of the phion and the chion correspondingly, 
 $\delta g$ is a counter-term of the renormalizarion of coupling, and
$$
L(p^2, \mu^2, m^2)=\int \frac{d^dq}{(2\pi)^d}\,\frac{1}{\mu^2+(p-q)^2}\,\frac{1}{m^2+q^2},
$$
$$
\Lambda(p_x, p_y)=\int \frac{d^dq}{(2\pi)^d}\,\frac{1}{\mu^2+q^2}\,\frac{1}{m^2+(p_x+q)^2}
\,\frac{1}{m^2+(p_y-q)^2},
$$

In the dimensional regularization
($d=6-\epsilon$):
$$
L(p^2, \mu^2, m^2)=
-\frac{\kappa^{-\epsilon}p^2}{192\pi^3\epsilon}-\frac{\kappa^{-\epsilon}(\mu^2+m^2)}{64\pi^3\epsilon}
+O(\epsilon^0),
$$
$$
\Lambda(0, 0) = \frac{\kappa^{-\epsilon}}{64\pi^3\epsilon}+O(\epsilon^0).
$$
Here $\kappa$ is a 't Hooft scale. We define the dimensionless coupling as
$$
g\rightarrow g_\epsilon=\kappa^{\epsilon/2}g,
$$
and by adopting the $MS$ scheme \cite{Collins},  we get the counter-terms:
\begin{equation}
\delta m^2=-\frac{g^2(\mu^2+m^2)}{64\pi^3\epsilon},\;
\delta \mu^2=-\frac{g^2 m^2}{32\pi^3\epsilon},\;
\delta z_1=\delta z_2 = -\frac{g^2}{192\pi^3\epsilon},\; 
\delta g = -\frac{g^3
\kappa^{\epsilon/2}}{64\pi^3\epsilon}.
\label{countert}
\end{equation}

\subsection{Renormalization Group. Hermitian theory}

The independence of initial (bare) quantities and unrenormalized Green functions from
the 't Hooft scale $\kappa$ leads to the renormalization group equation: 
\begin{equation}
 \Big(\kappa\frac{\partial}{\partial \kappa}+\beta\frac{\partial}{\partial g}
-m^2\gamma_m \frac{\partial}{\partial m^2}
-\mu^2\gamma_\mu \frac{\partial}{\partial \mu^2}-\frac{n}{2}\,\gamma_1-\frac{l}{2}\,\gamma_2\Big)\Gamma^{nl}=0.
\end{equation}
Here $\Gamma^{nl}$ is the one-particle-irreducible function with $n$ phion and $l$ chion tails.

Counter-terms (\ref{countert}) allow us to calculate  renormalization-group coefficients 
\footnote{We use  the notations of Collins
 \cite{Collins}. Note, that the complete renormalization-group analysis assumes also
an addition the linear term  $h\chi$ in Lagrangian (\ref{Lagr}) and the corresponding 
counter-term. We omit this simple generalization of calculations as non-essential  for our
consideration.}
\begin{equation}
\beta=\kappa\frac{d g}{d\kappa}=-\frac{\epsilon}{2}g-\frac{g^3}{256\pi^3}+O(g^5),
\label{beta} 
\end{equation}
\begin{equation}
 m^2\gamma_m=-\kappa\frac{d m^2}{d\kappa}=
\frac{g^2}{192\pi^3}\Big(2m^2+3\mu^2\Big)+O(g^4),
\label{gamma_m}
\end{equation}
\begin{equation}
 \mu^2\gamma_\mu=-\kappa\frac{d \mu^2}{d\kappa}=
\frac{g^2}{192\pi^3}\Big(6m^2 -\mu^2\Big)+O(g^4),
\label{gamma_mu}
\end{equation}
\begin{equation}
 \gamma_1=\kappa\frac{d \ln z_1}{d\kappa}=\gamma_2=\kappa\frac{d \ln z_2}{d\kappa}
= \frac{g^2}{192 \pi^3 }+O(g^4). 
\label{gamma}                                                                               
\end{equation}
These renormalization-group coefficients quite similar to corresponding coefficients
of  $\phi^3_6$--theory (see  \cite{Bender12}, \cite{Collins}). As for $\phi^3_6$--theory
the scalar Yukawa model near   $d=6$ possesses  only a Gaussian fixed point $g_*=0$, and 
near this point the couplings   scale according their scaling dimension.

At  $d=6$ the scalar Yukawa model is asymptotically free as the  $\phi^3_6$--theory.
The running coupling ( invariant charge) $\bar{g}(t, g)$ is a solution of equation
\begin{equation}
\frac{d \bar{g}}{dt}=\beta(\bar{g})=-\frac{\bar{g}^3}{256\pi^3}
\end{equation}
with the boundary condition
$
\bar{g}(t=0, g)=g.
$ Here $t =\ln \frac{p^2}{p_0^2}$.
 
For $\beta$--function (\ref{beta}) the solution of this equation at $d=6$ is
\begin{equation}
 \bar{g}^2=\frac{g^2}{1+\frac{g^2}{128\pi^3}\,t},
\label{g_bar}
\end{equation}
i.e. the model possesses the typical asymptotically-free behavior at high momenta 
with all consequences.

\subsection{Renormalization group. Non-Hermitian theory}
For the non-hermirian $PT$--symmetric theory one should make the substitution 
 $$g\rightarrow ig,\; \delta g\rightarrow i\delta g$$ in all formulae of above Subsections.
Thus, the expression for $\beta$--function takes the form
\begin{equation}
\beta=\kappa\frac{d g}{d\kappa}=-\frac{\epsilon}{2}g+\frac{g^3}{256\pi^3}+O(g^5)
\label{beta_i} 
\end{equation}
etc.

The situation in this case is also similar to  $\phi^3$--theory (see \cite{Bender12}).
$\beta$--function  vanishes, except of the Gaussian point $g _* = 0$,  
at the fixed point of Wilson-Fisher type:
\begin{equation}
g^2_*=128\pi^3\epsilon
\label{fixpoint} 
\end{equation}

Near   the Gaussian point couplings are still defined by their canonical dimensions.
Near the non-Gaussian fixed point (\ref{fixpoint}) the scale behavior is modified
in accordance with the linearized
renormalization group equations. At $d = 6$ fixed points merge into one  Gauss point.

The running coupling in this case is
\begin{equation}
 \bar{g}^2=\frac{g^2}{1-\frac{g^2}{128\pi^3}\,t},
\label{g_bar_i}
\end{equation}
i.e., the theory at large momenta has the  trivial-type  behavior, and  the
perturbation theory in this asymptotic region cannot be applicable.

\section{Beyond the perturbation theory}

\subsection{Shcwinger--Dyson equations}

To construct the non-perturbative approximation  we will use the formalism of 
Schwinger-Dyson equations (SDE).

The generating functional of  Green functions (vacuum
averages) of the model with Lagrangian  (\ref{Lagr})  is the functional integral
\begin{equation}
G(\eta, j)=\int
D(\phi,\phi^*,\chi)\exp\bigg\{\int dx\, {\cal L}(x)-
\int dxdy\,\phi^*(y)\eta(y, x)\phi(x)+\int dx \,j(x)\chi(x)
\bigg\}.
\end{equation}
Here $\eta$ is the bilocal source of phions
\footnote{A formalism
  of the bilocal source 
 was first
elaborated in the quantum field theory 
by Dahmen and Jona-Lasinio \cite{Dahmen}.
 We consider this using presumably as a convenient choice
of the functional variable.},
 $j$ is the single source of chions.\\
 The translational invariance of the functional integration measure leads to relations
$$
\int D(\phi,\phi^*, \chi) \frac{\delta}{\delta\phi^*(x)}\phi^*(y)
\exp\bigg\{\int dx {\cal L}(x)-\int dxdy
\phi^*(x)\eta(x,y)\phi(y)+\int dz j(z)\chi(z)\bigg\}=0,
$$
and
$$
\int D(\phi,\phi^*, \chi) \frac{\delta}{\delta\chi(z)}
\exp\bigg\{\int dx {\cal L}(x)-\int dxdy
\phi^*(x)\eta(x,y)\phi(y)+\int dz j(z)\chi(z)\bigg\}=0,
$$
which can be rewritten as the functional-differential
SDE for generating functional $G$:
\begin{equation}
g\frac{\delta^2G}{\delta\eta(y,x)\delta
j(x)}=(m_0^2-\partial_x^2)\frac{\delta G}{\delta\eta(y,x)}+
\int dx_1\eta(x,x_1)\frac{\delta
G}{\delta\eta(y,x_1)}+\delta(x-y)G
\label{SDE1}
\end{equation}
and
\begin{equation}
 g \frac{\delta
G}{\delta\eta(z,z)}+(\mu^2_0-\partial^2) \frac{\delta G}{\delta
j(z)}=j(z)G
\label{SDE2}
\end{equation}
Here $m_0$ and $\mu_0$ are bare phion and chion masses. 
Equation  (\ref{SDE2}) allows us to express
all  Green functions with chion legs in terms of functions that contain phions
only. For logarithm $Z=\log G$ this equation has the form
\begin{equation}
\frac{\delta Z}{\delta j(x)}=\int dx_1\,
D_c(x-x_1)\, j(x_1)-\int dx_1\,gD_c(x-x_1)\,
\frac{\delta Z}{\delta \eta(x_1,x_1)}.
\label{SDE2Z}
\end{equation}
(Here
$
D_c\equiv(\mu^2_0-\partial^2)^{-1}.
$)

 The differentiation of (\ref{SDE2Z}) over $\eta$ gives us
the three-point chion-phion function
\begin{equation}
V(xy|z)\equiv -\frac{\delta^2Z}{\delta j(z)\delta\eta(yx)}\bigg|_{\eta=j=0}=
\int dz_1\,g
D_c(z-z_1)\, Z_2\left(\begin{array}{cc}z_1&z_1\\x&y\end{array}\right),
\label{3point}
\end{equation}
where
\begin{equation}
 Z_2\left(\begin{array}{cc}x&y\\x'&y'\end{array}\right)
\equiv\frac{\delta^2 Z}{\delta\eta(y',x')\delta\eta(y, x)}\,\bigg|_{\eta=j=0}
\end{equation}
is the two-particle phion function.
The differentiation of  (\ref{SDE2Z}) over  $j$ with
taking into account equation (\ref{3point}) gives us the chion
propagator:
\begin{equation}
 D(x-y)\equiv\frac{\delta^2 Z}{\delta j(y)\delta
j(x)}\bigg\vert_{\eta=j=0}
=D_c(x-y)+\int dx_1dy_1\,g^2
D_c(x-x_1)Z_2\left(\begin{array}{cc}x_1& x_1\\y_1&y_1\end{array}\right)
D_c(y_1-y)
\label{propchi}
\end{equation}
etc.
Thus, for a complete description of the model we need to know
phion Green function only. 

Excluding with  the help of the SDE (\ref{SDE2}),
 a differentiation over $j$ in SDE (\ref{SDE1}),
we obtain at $j=0$ the equation
\begin{eqnarray}
\int dx_1\,g^2
D_c(x-x_1)\frac{\delta^2 G}{\delta\eta(x_1,x_1)\delta\eta(y,x)}
+(m^2_0-\partial_x^2)\frac{\delta G}{\delta\eta(y,x)}+
\nonumber \\
+\int dy_1\eta(x,y_1)
\frac{\delta G}{\delta\eta(y,y_1)}+\delta(x-y)G=0,
\label{SDE1G}
\end{eqnarray}
which only  contains
the derivatives over the bilocal
source $\eta$.

Since $
\delta^2 G/\delta\eta(y', x')\delta\eta(y, x)=
<\phi(x)\phi^*(y)\phi(x')\phi^*(y')>$, then
Bose-symmetry entails the relation 
$$
\frac{\delta^2 G}{\delta\eta(y', x')\delta\eta(y, x)}=
\frac{\delta^2 G}{\delta\eta(y', x)\delta\eta(y, x')},$$
reflecting  crossing symmetry of the two-particle  function,
and, accordingly, the equation (\ref{SDE1G}) can be written
as
\begin{eqnarray}
\int dx_1\,g^2
D_c(x-x_1)\frac{\delta^2 G}{\delta\eta(x_1,x)\delta\eta(y,x_1)}
+(m^2_0-\partial_x^2)\frac{\delta G}{\delta\eta(y,x)}+
\nonumber \\
+\int dy_1\eta(x,y_1)
\frac{\delta G}{\delta\eta(y,y_1)}+\delta(x-y)G=0.
\label{SDE1cross}
\end{eqnarray}
Both equations give the same  coupling-constant perturbation series, and are
completely equivalent from the point of view of some visionary exact solutions of
Schwinger-Dyson equations. However, these equations give different non-perturbative expansion.
This is due to the incomplete structure of the leading-order multi-particle functions of such expansions
  in terms of crossing symmetry.
It is a peculiar feature of some non-perturbative approximations.
In order to restore crossing symmetry lost in the leading-order approximation, it is necessary to consider
the next-to-leading-order approximation. 
(A more detailed discussion of this issue see in the papers \cite{Rochev13}, \cite{Rochev15}
and references  therein).

Equation  (\ref{SDE1G}) can be used  for the construction of the 
mean-field expansion (see \cite{Rochev13}). In the language of Feynman diagrams the
leading order of this expansion 
corresponds to the summation of the chains and its structure  actually reproduce the
renormalization-group summation of the previous section.

In this paper we consider the expansion,  based on the equation  (\ref{SDE1cross})
(see also \cite{Rochev15}).
In the language of Feynman diagrams the leading order of
this expansion corresponds to the summation of ladder graphs, so we'll call it the ladder
expansion.

For logarithm $Z=\log G$ this equation has the form
\begin{eqnarray}
\int dx_1\,g^2
D_c(x-x_1)\bigg[\frac{\delta^2 Z}{\delta\eta(x_1,x)\delta\eta(y,x_1)}
+\frac{\delta Z}{\delta \eta(x_1,x)}\,\frac{\delta Z}{\delta \eta(y,x_1)}\bigg]+
\nonumber \\
+(m^2_0-\partial_x^2)\frac{\delta Z}{\delta\eta(y,x)}+
\int dy_1\eta(x,y_1)
\frac{\delta Z}{\delta\eta(y,y_1)}+\delta(x-y)=0.
\label{SDE1Z}
\end{eqnarray}

\subsection{Legendre transform}
Equation
\begin{equation}
\frac{\delta Z}{\delta \eta(y,x)}=-\Delta(x,y|\eta), 
\label{prophi}
\end{equation}
which determines the phion propagator  can be regarded as an equation that determines
implicitly $\eta$ as a functional of $\Delta$:
$$
\eta=\eta[\Delta].
$$
Assuming  the unique solvability of the equation (\ref{prophi}), we can
move to a new function variable $\Delta$ and define the
generating functional of Legendre transform
\begin{equation}
\Gamma[\Delta]
=Z+\int dxdy \,\Delta(x, y)\, \eta (y, x).
\label{GammaL}
\end{equation}
From definitions (\ref{prophi}) and (\ref{GammaL}) it follows that
\begin{equation}
\frac{\delta \Gamma}{\delta \Delta(y,x)}=\eta(x, y|\Delta),
\end{equation}
and SDE  (\ref{SDE1Z}) takes the form
\begin{eqnarray}
\frac{\delta \Gamma}{\delta \Delta(y,x)}=\Delta^{-1}(x,y)-
(m_0^2-\partial^2)\delta(x-y)+g^2D_c(x-y)\Delta(x,y)+
\nonumber \\
+\int dx_1xy_1\,g^2 D_c(x-x_1)\frac{\delta^2 Z}{\delta \eta(x,x_1)\delta\eta(y_1,x_1)}\,
\Delta^{-1}(y_1,y).
\label{SDEGamma} 
\end{eqnarray}
In this equation, it is assumed that $\delta^2Z/\delta\eta^2$  is a functional of 
new functional variable $\Delta$, what can be done, using the condition of
connection
\begin{equation}
\int dx_1dy_1\,\frac{\delta^2 \Gamma}{\delta \Delta(y,x)\delta\Delta(y_1,x_1)}\,
\frac{\delta^2 Z}{\delta \eta(y',x')\delta\eta(x_1,y_1)}\,
=-\delta(x-y')\delta(x'-y),
\label{connection}
\end{equation}
which follows from the relation
$$
\frac{\delta \eta(x,y)}{\delta \eta(y',x')}=\delta(x-y')\delta(x'-y).
$$

\subsection{Ladder expansion}

SDE (\ref{SDEGamma}) tells us a non-perturbative expansion of  the generating functional
 $\Gamma=\Gamma_0+\Gamma_1+\cdots$, which based on the following leading approximation
\begin{equation}
\frac{\delta \Gamma_0}{\delta \Delta(y,x)}=\Delta^{-1}(x,y)-
(m_0^2-\partial^2)\delta(x-y)+g^2D_c(x-y)\Delta(x,y).
\label{Gamma0} 
\end{equation}
Next-to-the-leading-order equation is
\begin{equation}
\frac{\delta \Gamma_1}{\delta \Delta(y,x)}=
\int dx_1dy_1 g^2D_c(x-x_1)\,\frac{\delta^2 Z_0}{\delta \eta(x_1,x)\delta\eta(y_1,x_1)}\,
\Delta^{-1}(y_1,y).
\end{equation}
where  $\delta Z_0/\delta \eta^2$ is a functional of  $\Delta$, defined by condition of
connection
(\ref{connection}).

At the  source being switched off, equation (\ref{Gamma0}) is the equation for the 
leading-order phion
propagator:
\begin{equation}
\Delta_0^{-1}(x-y)=(m_0^2-\partial_x^2)\delta(x-y)
-g^2D_c(x-y)\Delta_0(x-y).
\label{Delta0}
\end{equation}

A differentiation of equation (\ref{Gamma0}) on $\Delta$ and taking into account 
 connection condition
(\ref{connection}) together with equation (\ref{3point}) gives us the equation for the three-point
function:
\begin{equation}
\int dx_1dy_1\, \Delta^{-1}_0(x,x_1)V_0(x_1,y_1|z_1)\Delta^{-1}_0(y_1,y)=
g^2D_c(x-z)\delta(x-y)+g^2D_c(x-y)V_0(x,y|z).
\label{3point0}
\end{equation}

\subsection{Phion propagator}

Lets go to the equation  (\ref{Delta0}) for the phion propagator.
To eliminate ultraviolet divergences in  equation (\ref{Delta0}) 
is sufficient to introduce counter-terms of phion-field renormalization
 $z_1$  and mass $\delta m^2$.
The normalization of the renormalized propagator $\Delta (p^2) $
at zero momentum
$$
\Delta^{-1}(p^2=0)=m^2, \;
\frac{d \Delta^{-1}}{dp^2}\bigg|_{p^2=0}=1
$$
leads to the  renormalized equation in momentum space
\begin{equation}
\Delta^{-1}(p^2)=m^2+p^2+\Sigma_r(p^2),
\label{Delta}
\end{equation}
where $\Sigma_r(p^2)=\Sigma(p^2)-\Sigma(0)-p^2\Sigma'(0)$ is the renormalized
mass operator, and 
$$
\Sigma(p^2)=-g^2\int \frac{d^6q}{(2\pi)^6}\;D_c(p-q)\Delta(p).
$$
Below we consider the case of massless chion: $D_c=1/k^2$. In this case
nonlinear integral equation (\ref{Delta}) can be reduced to an integral
Volterra-type equation, which, in turn, is reduced to a differential equation.
Using the formula massless integration in six-dimensional space
\begin{equation}
\int \frac{d^6q}{(2\pi)^6}\,\frac{\Phi(q^2)}{(p-q)^2}=
\frac{1}{128\pi^3}\bigg[\int_0^{p^2}q^2\Phi(q^2)dq^2\bigg(\frac{q^2}{p^2}-
\frac{1}{3}\Big(\frac{q^2}{p^2}\Big)^2
\bigg)+\int_{p^2}^\infty q^2\Phi(q^2)dq^2\bigg(1-
\frac{1}{3}\,\frac{p^2}{q^2}\bigg)\bigg]
\label{MLI}
\end{equation}
we obtain
\begin{equation}
 \Sigma_r=
-\frac{g^2\,p^2}{384\pi^3}
\int_0^{p^2} dq^2\Delta(q^2)\bigg(1-\frac{q^2}{p^2}\bigg)^3.
\label{sigma}
\end{equation}
Introducing dimension-less function  $$u(t)=\frac{1}{m^2}\Delta^{-1},$$ where 
$
t=\frac{p^2}{m^2},
$
we obtain the integral equation 
\begin{equation}
 u(t)=1+t-\frac{g^2}{384\pi^3t^2}\int_0^t\frac{dt'}{u(t')}(t-t')^3,
\label{u_int}
\end{equation}
which is reduced to the non-linear fourth-order differential equation
\begin{equation}
 (t^2u)''''=-\frac{g^2}{64\pi^3u}.
\label{u_diff}
\end{equation}
This differential equation enables us to calculate the asymptotics
of   $u(t)$ for large $t$:
\begin{equation}
 u(t)\simeq At\sqrt{\log t},
\label{asymp_u}
\end{equation}
where
$$
A^2=-\frac{g^2}{192\pi^3},
$$
i.e., for Hermitian theory with $g^2> 0$ the asymptotic behavior becomes purely imaginary. In order to
understand what is happening with the propagator in  Euclidean region,  consider a simplified model
with the same UV behavior. This model is based on the following approximation of
 mass operator (\ref{sigma}) in a high-momentum region:  
\begin{equation}
 \Sigma_r=
-\frac{g^2\,p^2}{384\pi^3}
\int_0^{p^2} dq^2\Delta(q^2)\bigg(1-\frac{q^2}{p^2}\bigg)^3\approx
-\frac{g^2\,p^2}{384\pi^3}
\int_0^{p^2} dq^2\Delta(q^2).
\label{sigma_approx} 
\end{equation}
The equation  for the inverse propagator  $u$  takes the form:
\begin{equation}
 u(t)=t-\frac{g^2\,t}{384\pi^3}\int_{t_0}^t\frac{dt'}{u(t')}.
\label{u_approx}
\end{equation}
The cutoff  at the lower limit of integration  is introduced in order  to avoid mass
singularities (in the case insignificant).

The exact solution of equation (\ref{u_approx}) is
\begin{equation}
 u(t)=t\,\sqrt{1-\frac{g^2}{192\pi^3}\,\log\frac{t}{t_0}},
\end{equation}
i.e., an asymptotic behavior at large momentum given by the same formula
(\ref{asymp_u}).

Thus, we can conclude that for the usual Hermitian theory with $g^2> 0$
the propagator in the ladder approximation has the non-physical non-isolated singularity
  in the Euclidean region, while
for the non-Hermitian theory with $g^2 <0$, this singularity moves in a pseudo-Euclidean region, i.e.,
the non-Hermitian theory is more preferable from the standpoint of the analytical properties of the propagator.

\section{Conclusion}

Our results demonstrate, that the non-Hermitian $PT$--symmetric scalar Yukawa model has 
 interesting properties both perturbative and non-perturbative.
In the perturbation region of small momenta, $ig(\phi^*\phi\chi)_6 $ theory similar in their
properties to Hermitian $g(\phi^*\phi)^2_4$  theory, i.e., energetically stable and has,
in addition to the Gaussian fixed point, a non-trivial fixed point of  Wilson-Fisher type.
As expected, the properties of the scalar Yukawa model in the perturbative region
completely analogous to the corresponding properties of $ig\phi^3_6$ theory (see \cite{Bender12}).
The non-perturbative ladder expansion of  Section 3 reveals new
 interesting feature of the non-Hermitian model. While in the Hermitian version of theory the 
phion propagator 
 has the non-physical non-isolated singularity in the  Euclidean region of momenta, the non-Hermitian theory  
substantially free of this drawback, as the singulatity  moves to the pseudo-Euclidean
region.

For a complete description of the leading-order ladder expansion,
 including its renormalization group analysis, it is necessary to
solve  equation  (\ref{3point0}) for the three-point function. This is a very difficult task, since
this equation contains a nontrivial phion propagator, described by  equation (\ref{Delta0}). Perhaps for
of the renormalization-group analysis and
clarify the nature of the behavior of couplings in the asymptotic region is sufficient to solve a more limited
 problem, namely the calculation of the vertex function at zero momentum (which, however, also
very difficult). We can assume that in the Hermitian Theory  the theory retain the  property
of asymptotic
freedom, and everything will return to own (if the effects of
the energy  unstability of the   Hermitian model not appear). For the non-Hermitian $PT$--symmetric theory 
 a prediction of 
the answer is harder. In any case, the results indicate that the non-Hermitian
scalar Yukawa model has, compared with the Hermitian version, a number of attractive features,
that make it a very interesting object of study.

\section*{Aknowlegements}
Author is grateful to the participants of IHEP Theory Division Seminar
for useful discussion.

\end{document}